\documentclass{article}

\usepackage[preprint]{neurips_2026}

\usepackage[utf8]{inputenc} 
\usepackage[T1]{fontenc}    
\usepackage{hyperref}       
\usepackage{url}            
\usepackage{booktabs}       
\usepackage{amsfonts}       
\usepackage{nicefrac}       
\usepackage{microtype}      
\usepackage{subcaption}
\usepackage{graphicx}
\usepackage{amsmath}
\usepackage{tikz}
\usepackage{inconsolata}
\usepackage{amssymb}  
\usepackage{makecell}
\usepackage{comment}
\usepackage[textsize=tiny]{todonotes}
\usepackage[most]{tcolorbox}
\usepackage{siunitx}
\usepackage[capitalize]{cleveref}
\usepackage{xcolor}
\usepackage{placeins}
\definecolor{darkgreen}{RGB}{0, 120, 0}
\definecolor{darkred}{RGB}{160, 0, 0}
\definecolor{promptblue}{RGB}{20, 50, 120}

\usepackage{multirow}
\usepackage{pgf}
\usepackage{import}

\everymath=\expandafter{\the\everymath\displaystyle}

\usepackage{enumitem}
\usepackage{wrapfig}
\usepackage{hyperref}
\definecolor{brown}{RGB}{90, 0, 0}
\hypersetup{
    colorlinks=true,
    citecolor=brown,    
    linkcolor=brown,    
    urlcolor=brown,     
}

\newcommand{\prompt}[1]{{\small\ttfamily\color{promptblue}#1}}

\usepackage[most]{tcolorbox}

\newtcolorbox{promptboxenv}{
    colback=white,
    colframe=gray!70,
    boxrule=0.5pt,
    arc=0pt,
    left=6pt, right=6pt, top=2pt, bottom=2pt,
    enhanced,
    breakable,
    before skip=6pt,
    after skip=10pt,
}

\newcommand{\promptex}[1]{%
    \begin{promptboxenv}
    \textit{Example:} \prompt{#1}
    \end{promptboxenv}%
}

\usepackage{pifont}
\newcommand{\xmark}{{\color{darkred}\ding{55}}}
\newcommand{\cmark}{{\color{darkgreen}\ding{51}}}

\title{Fingerprinting Inference Systems of\\ Large Language Models}

\author{%
  Anna Wimbauer \\
  BIFOLD \& TU Berlin \\
   \And
   Jonas Möller \\
  BIFOLD \& TU Berlin \\
    \AND
   Erik Imgrund \\
  BIFOLD \& TU Berlin \\
  \And
  Konrad Rieck \\
  BIFOLD \& TU Berlin \\
}

\begin{document}

\maketitle

\begin{abstract}

The behavior of LLMs does not depend solely on the model itself. Components of the inference system, such as the inference engine, attention backend, and hardware platform, subtly influence how inputs are processed. 
These components differ in their implementations and thereby induce small numerical deviations across systems when running the same model.
While prior work has established the theoretical existence of such deviations, their security implications have remained unexplored.
In this paper, we show that these deviations are characteristic of specific components and propagate to observable textual outputs, exposing the inference system to any party that can query the model.
Building on this observation, we introduce a fingerprinting method that analyzes the prompt-response behavior of LLMs to identify components of the inference system. Our empirical evaluation demonstrates that the inference engine, attention backend, and underlying hardware platform can be identified reliably, even when the LLM is operated at non-zero temperature.
We show that preventing fingerprinting is fundamentally hard, as it would require eliminating numerical differences between hardware and software stacks. We therefore propose partial mitigations and discuss their impact.

\end{abstract}
\section{Introduction}

Efficient inference of LLMs requires carefully engineered systems composed of specialized software and hardware components. 
While the underlying mathematics is the same, the actual computation in these systems is shaped by the selected components and their low-level implementation details. 
As a result, the same input produces slightly different numerical results across systems, due to the non-associativity of floating-point arithmetic~\citep{schlogl2024causes, yuan2025understanding}. 
For example, running an LLM on an Nvidia A100 versus an H100 GPU produces tiny changes in the computed logits, typically invisible to the user. Such deviations are generally treated as harmless artifacts that merely hinder reproducibility.

In this paper, we challenge this assumption and show that numerical deviations leak information about the inference system of an LLM, including the inference engine, attention backend, and hardware platform. Such leakage has direct security and trust implications. An adversary who knows the engine in use can target recent critical vulnerabilities in widely deployed implementations like vLLM and SGLang~\citep{wu2026promptpeek, wu2026cache,cve_2026_22778,cve_2026_27893,cve_2026_5760}. Conversely, a customer who can identify the underlying GPU can verify whether a service actually provides the hardware it promises.

Uncovering a signal from the numerical deviations of an LLM inference system, however, is non-trivial. While prior work~\citep{zhang2024hardware} has shown how information can be extracted from model logits, common LLM deployments do not expose logits, and only the input and output tokens are available for analysis.
We address this challenge by introducing a \emph{fingerprinting method} for LLM inference systems. Our method uses specifically crafted prompts that induce numerical deviations at distinctive computational stages of inference, such as autoregressive decoding, prompt processing (prefill), and caching. We encode the resulting prompt-response pairs into a joint feature representation and train a classifier to identify the system components in use.

We empirically evaluate our approach on systems built from a range of components (4 inference engines, 6 attention mechanisms, and 3 hardware devices). At temperature zero, our method identifies each component perfectly across different LLMs. At non-zero temperatures, where sampling introduces stochasticity, our method identifies components with 76--80\% accuracy in many configurations. We further examine the robustness of our approach under external factors and find that component fingerprinting remains feasible across a range of realistic settings.

Unfortunately, defending against this type of fingerprinting is non-trivial. Unifying the computation of inference systems would eliminate the vulnerability but forfeit the advantages of diverse software and hardware. Similarly, adding noise to the output would mask the leak at the cost of applications that require deterministic decoding. We later explore these tradeoffs and partial mitigations.

In summary, we make the following contributions:
\begin{itemize}[leftmargin=*]
\item \textbf{Fingerprinting of LLM inference systems.} We introduce the first fingerprinting method that identifies inference engines, attention backends, and hardware platforms from prompt-response pairs alone, without access to logits.

\item \textbf{Evaluation across realistic configurations.} We evaluate our approach on systems built from 4 inference engines, 6 attention mechanisms, and 3 hardware devices, achieving perfect identification at temperature zero and reliable identification under sampling.

\item \textbf{Fingerprint robustness and defenses.} We show that fingerprinting remains effective under realistic deployment variations, and thus designing robust defenses becomes difficult. As a remedy, we discuss mitigations that trade off properties such as determinism and utility.

\end{itemize}

\section{Background}
\label{sec:background}

We begin by introducing the two building blocks of our fingerprinting approach: the ecosystem of inference systems, and the numerical deviations that their components induce during computation. 

\paragraph{Inference systems.}
\begin{wrapfigure}[13]{r}{0.47\textwidth}
\centering
\vspace{-2mm}
\includegraphics[width=0.47\textwidth]{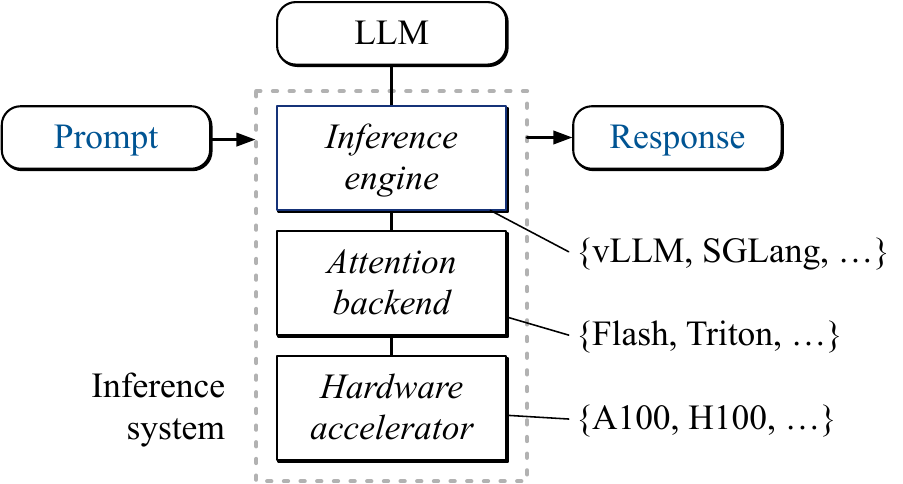}\\[0.9mm]
\caption{An LLM inference system.}
\label{fig:inference-system-overview}
\end{wrapfigure}
An LLM inference system comprises several interacting layers, as shown in \autoref{fig:inference-system-overview}. At the lowest level, hardware accelerators such as Nvidia GPUs provide the high-throughput computation that dominates inference cost. Above the hardware, a backend implements an attention mechanism~\citep{vaswani2017attention} together with optimizations such as tiling, kernel fusion, and IO-aware scheduling. At the top, an inference engine such as vLLM or SGLang orchestrates execution: it manages request batching, key-value caches, and multi-device parallelism while exposing a clean interface to the user.

From the user's perspective, the system behaves like a black box that maps a prompt to a response, with the underlying stack hidden from view. Because components differ in their implementations and resource usage, minor deviations in the underlying computations propagate through this stack, though they typically remain imperceptible in the response.

\paragraph{Numerical deviations.}
The root cause of these numerical deviations is the non-associative nature of floating-point arithmetic:
due to limited precision, the order in which operations are executed affects the result. That is, whenever a component reshapes a tensor, alters a reduction order, or schedules computations differently from an alternative implementation, the resulting calculations begin to diverge. 
\citet{yuan2025understanding} demonstrate this effect at the system level, while \citet{golden2024flashattentionstable} show that attention backends exhibit implementation-dependent deviations. These small differences are a notorious source of reproducibility issues, yet have been considered harmless so far.

We can illustrate these deviations with a simple example. Consider two
matrices $A, B \in \mathbb{R}^{100 \times 100}$ with all entries of $A$
equal to $0.02$ and all entries of $B$ equal to $0.005$. Mathematically,
we now have $\operatorname{tr}(A^\top B) = 1,$
yet evaluating this quantity on different GPUs in single precision yields
\begin{align*}
  \operatorname{tr}(A^\top B) &\approx 0.999999%
  \textcolor{darkred}{9403953552} \quad \text{(Nvidia A100)}, \\
  \operatorname{tr}(A^\top B) &\approx 0.999999%
  \textcolor{darkred}{0463256836} \quad \text{(Nvidia H100)}.
\end{align*}
For an LLM with billions of parameters and many such matrix operations per forward pass, these per-operation deviations compound, so that even a short question can yield diverging outputs depending on the underlying inference system, as shown in \autoref{tab:deviate_example}. Note that in this example, the logits differ across all three configurations, yet the predicted token flips for only one of them.


\begin{table}[htbp]
\centering\small
\caption{Response logits for the same model across inference systems for the prompt ``\prompt{Did England lose a game in the 1966 World Cup?}''. Greedy decoding (temperature $0$) is used so that any deviation arises from the components of the systems.} 
\label{tab:deviate_example}
\vspace{2mm}
\begin{tabular}{lllc S[table-format=1.4, table-text-alignment=center] S[table-format=1.4, table-text-alignment=center]}
\toprule
& & & & \multicolumn{2}{c}{\textbf{Response Probabilities}} \\
\cmidrule(lr){5-6}
\textbf{Model} & \textbf{Engine} & \textbf{Attention} & \textbf{GPU} & {\textbf{Yes}} & {\textbf{No}} \\
\midrule
Qwen2.5-7B-Instruct & SGLang & FlashAttention & A100 & \color{darkred} 0.3775 & \color{darkgreen} 0.6224  \\ 
Qwen2.5-7B-Instruct & SGLang & FlashInfer & H100 & \color{darkred} 0.4999 & \color{darkgreen} 0.4999 \\ 
Qwen2.5-7B-Instruct & TensoRT-LLM  & FlashInfer & H100 & \color{darkgreen} 0.5622 & \color{darkred} 0.4378 \\ 
\bottomrule
\end{tabular}
\end{table}


\section{Fingerprinting Approach}
\label{sec:method}

We now introduce a fingerprinting method that identifies system components from prompt–response pairs. Specifically, we aim to infer the inference engine, attention backend, and hardware platform, which constitute the core components described in Section~\ref{sec:background}.
Our approach proceeds in two stages. First, we construct prompt sets that target distinct computational phases of the inference stack. Second, we derive joint embeddings from the resulting responses and train a classifier to identify instances of the considered system components.

\paragraph{Threat model.}
We consider an attacker who aims to identify the components of a deployed LLM inference system. Our attack assumes knowledge of the deployed model and system prompt, both of which can be recovered through existing attack such as model fingerprinting~\citep{Pasquini2025LLMmap, gubri2024trap} and prompt extraction~\citep{hui2024pleak, yang2025prsa}. The attacker has only chat access: they can submit arbitrary prompts and observe the resulting text responses. Log-probabilities or internal states are inaccessible. For preparation, we further assume offline access to a set of reference systems on which training data can be collected (Section~\ref{sec:experiments}).

\subsection{Prompt Set Construction}
\label{sec:prompts}

When a prompt is processed by an inference system, it traverses a series of computational stages before the response is obtained. These stages depend on the prompt's length and structure, and typically include prefill, chunked prefill, KV-cache reuse, and autoregressive decoding. Each stage exercises the inference stack in a different way and thus reveals different numerical deviations. Since no single prompt can activate all stages, we construct four prompt sets $\mathcal{P}_1, \ldots, \mathcal{P}_4$, each tailored to a specific stage.

\begin{description}[leftmargin=2.2em, style=unboxed, itemsep=4pt]
   \item \emph{$\mathcal{P}_1$: Rare token prompts.}
    These prompts contain a rare token sequence, for example, a string of non-standard Unicode characters, inside unrelated text and ask the model to retrieve it. Because the target token sits close to other low-probability tokens, even minor deviations accumulated during \emph{autoregressive decoding} can shift the prediction across system configurations.

    \promptex{Remember this unique identifier: \detokenize{ﴰ}  Scientists have recently discovered new species deep in the Amazon rainforest. [$…$]
    What was the unique identifier mentioned at the very beginning? Reply with only the identifier itself.}

\item \emph{$\mathcal{P}_2$: Binary decision prompts.}
    These prompts are designed to elicit a single yes/no token, limiting
generation to one decoding step in non-reasoning models. This effectively
isolates the \emph{prefill stage}: since there is virtually no autoregressive
generation, any variation in the response must originate from how the prompt
was processed during prefill.

    \promptex{Is the following statement true? Answer only Yes or No.
    Did England lose a game in the 1966 World Cup?}

\item \emph{$\mathcal{P}_3$: Long-context prompts.}
    These prompts are long and ask the model to repeat back a specific number (non-rare tokens) embedded within. Due to their length, they trigger \emph{chunked prefill}, where long contexts are split across multiple prefill passes. Any deviation in the returned number then points to numerical divergences introduced by the chunking process.
    
    \promptex{Find the value of NUM in the following text. Reply with only the number, nothing else.  compatible mathematicalolan Las shrouded companionsmite StreamstallSel arcanePrin  [$…$] NUM=745.
    aspirationacha take amyJonathaniped Tak audit Stocksever Cour Membership}

\item \emph{$\mathcal{P}_4$: Repetition prompts.}
    These prompts ask the model to repeat a fixed sentence multiple times. This repetition triggers \emph{KV-cache prefix reuse}, where previously computed key-value states are loaded instead of recomputed. The observable signal is the number of times the sentence is actually repeated, indicating how cached states are managed in the system.

    \promptex{Repeat the following sentence 100 times: It was the best of times, it was the worst of times.}
    
\end{description}

Together, these prompt sets provide a comprehensive view of model behavior across common computational stages.
While this coverage is sufficient for fingerprinting the components considered in this work, we do not claim it to be exhaustive.
Other prompt designs may elicit further behavior within inference systems.

\subsection{Embeddings and Classification}
\label{sec:embeddings}

Given a prompt set $\mathcal{P}_i = \{p_1, \ldots, p_n\}$, we query the target system and collect the corresponding responses $\mathcal{R}_i = \{r_1, \ldots, r_n\}$. To represent these responses in a form suitable for classification, we associate each prompt set $\mathcal{P}_i$ with a scoring function $s_i : \Sigma^* \rightarrow [0,1]$ that captures the property of interest for that set. Applying $s_i$ to each response yields the per-set embedding vector
\begin{equation*}
\phi_i : (\Sigma^*)^n \rightarrow [0,1]^n, \qquad (r_1, \ldots, r_n) \mapsto \bigl(s_i(r_1), \ldots, s_i(r_n)\bigr).
\end{equation*}
For prompt sets $\mathcal{P}_1$ and $\mathcal{P}_3$, the scoring function $s_i$ is an indicator that returns $1$ if the response contains the expected tokens and $0$ otherwise. For $\mathcal{P}_2$, the function $s_2$ returns $1$ if the response starts with ``yes'' and $0$ otherwise. For $\mathcal{P}_4$, where the response is a repetition count, $s_4$ normalizes this count to $[0,1]$ relative to the requested number of repetitions. The resulting per-set embedding vectors $\phi_1, \ldots, \phi_4$ are concatenated into a single feature vector that serves as the input to the classifier.

\paragraph{Training.}
We construct training data by instantiating inference systems for all combinations of the considered components and querying each with the prompt sets defined above. The resulting prompt–response pairs are mapped to feature vectors via the embedding $\phi$. In our experiments (Section~\ref{sec:experiments}), this yields 30 systems. We train a separate random forest classifier for each target label: the inference engine, the attention backend, and the GPU type.

\paragraph{Fingerprinting.}
To fingerprint a target system, the attacker queries it with the prompt sets $\mathcal{P}$, constructs the feature vector via $\phi$, and obtains a prediction for each component from the corresponding classifier.
To accommodate systems operated at non-zero temperature, we collect $l$ prompt-response pairs per prompt during training and $k$ pairs during fingerprinting. The classifiers therefore learn to generalize across multiple samples per prompt, and we can aggregate predictions by majority voting when stochastic decoding is used.


\subsection{Practical Considerations}
\label{sec:practical_con}

Our approach involves three main parameters that the adversary must fix prior to launching the attack. The first is the set of candidate components considered for each target label. These must be enumerated in advance, as the adversary needs to instantiate a working reference system for every combination to collect training data. This is infeasible for proprietary systems such as those operated by OpenAI, Anthropic, or Google. However, it is straightforward for open-source deployments, where engines, attention backends, and hardware can be freely combined and reproduced. Our evaluation focuses on this open-source setting and uses widely adopted implementations.

The remaining two parameters control the cost of training and fingerprinting. The size $|\mathcal{P}|$ of the joined prompt set governs how much signal the classifier receives, and the number of repetitions $k$ controls robustness to non-deterministic decoding. Both can in principle be increased to improve accuracy, but at the price of additional queries to each system. We investigate both parameters in the following. 

\section{Evaluation}
\label{sec:experiments}

We continue with an empirical evaluation of our fingerprinting approach across different models and inference systems. In particular, we assess accuracy under deterministic and non-deterministic decoding, and investigate the role of prompt-set size and the number of repetitions. A detailed analysis of external factors is deferred to \cref{sec:experiments}.

\paragraph{Inference systems and models.}
For our evaluation, we consider inference systems spanning four popular inference engines, six attention backends, and three hardware accelerators (\autoref{tab:config_setup}). Consequently, random chance accuracy is $25\%$ for inference engine, $17\%$ for attention backend, and $33\%$ for GPU type.
These selected components are among the most widely used open-source implementations in their categories, reflecting realistic deployment scenarios.
Since not all backends are supported by all engines, the valid configurations form a subset of the Cartesian product, yielding $30$ distinct combinations. A detailed listing of these combinations is given in \autoref{sec:detailed_technical_setup}. 

As LLMs in these systems, we use three models compatible with all configurations: \emph{Qwen3-4B} (reasoning), \emph{Qwen2.5-7B-Instruct}, and \emph{Llama-3.2-3B-Instruct}. Model selection was constrained by cross-engine compatibility, as the most recent models are not yet fully supported across all engines; exact version strings are listed in \autoref{sec:detailed_technical_setup}. Although the parameter counts of these models are moderate, the effects of numerical deviations grow with the amount of computation. Our setup can therefore be regarded as a conservative setup for evaluating fingerprinting capabilities.

\newcommand{\mystar}{{\raisebox{0.1em}{\tiny\color{promptblue}$\bigstar$}}}

\begin{table}[bhp]
  \centering \small
  \caption{Considered components for inference systems. GitHub star counts (\mystar) are reported where the component is distributed as a standalone repository (May 2026).}
  \label{tab:config_setup}
  \vspace{3pt}
  \small
  \setlength{\tabcolsep}{4pt}
  \renewcommand{\arraystretch}{1.25}
  \begin{tabular}{@{}lcl@{}}
    \toprule
    \textbf{Component} & \textbf{\#} & \textbf{Configuration options} \\
    \midrule
    Inference Engine    & 4 & vLLM~(\mystar\,79k), SGLang~(\mystar\,27k), TensorRT-LLM~(\mystar\,13k), LMDeploy~(\mystar\,8k) \\
    Attention Backend   & 6 & FlashAttention~(\mystar\,24k), FlashInfer~(\mystar\,6k), Triton, TRT-LLM, PyTorch, TurboMind \\
    GPU Type  & 3 & Nvidia H100, Nvidia A100, Nvidia L4 \\
    \bottomrule
  \end{tabular}
\end{table}

\paragraph{Fingerprinting setup.}

We instantiate the four prompt sets from Section~\ref{sec:method} as follows: $\mathcal{P}_1$ contains 1700 natural text fragments with a rare token embedded mid-sequence. $\mathcal{P}_2$ consists of \num{5000} yes/no questions drawn from the BoolQ benchmark~\citep{clark2019boolq}. $\mathcal{P}_3$ comprises 150 prompts that embed a numeric identifier within long contexts of varying length. $\mathcal{P}_4$ contains 100 prompts, each instructing the model to repeat a distinct sentence or word $100$ times.

To obtain training and test instances, we query each of the 30 configurations with these prompts using the default system prompt of the respective model. 
Since batch size and batch position influence model outputs, all prompts are submitted at a fixed batch position across experiments. We evaluate the influence of batch size separately in \cref{sec:analysis}.

\subsection{Fingerprinting Systems with Deterministic Decoding}
\label{sec:results_temp0}

We begin with the case in which the inference system uses deterministic decoding, that is, operates at temperature zero. This is useful in many production deployments, including code-generation tools, structured-output pipelines, and tool-based agents, where reproducibility and output consistency are required~\citep{yuan2025understanding, renze-2024-effect}. This setting is favorable to the attacker, as any differences between systems are directly attributable to the inference  stack.

Under deterministic decoding, our fingerprinting method achieves perfect accuracy across all components and models (see $T=0$ in \autoref{tab:main_results}). Without sampling noise, the classification problem reduces to a retrieval task: the classifier merely needs to match prompt-response patterns to the corresponding system configuration in the training set.

\begin{figure}[h]
    \centering
    \scalebox{0.87}{\import{figures}{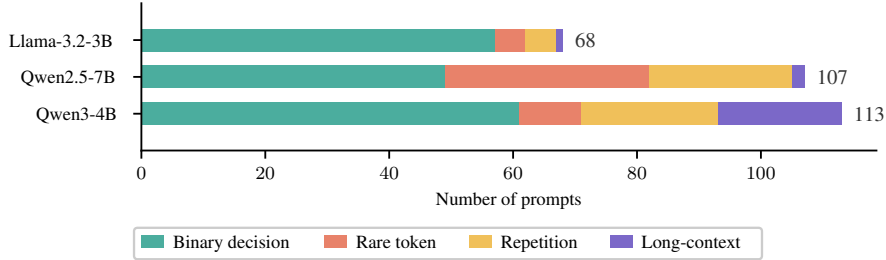}}
    \caption{Reduction of prompts for deterministic decoding.}
    \label{fig:prompt_distribution}
\end{figure}

Given this result, we ask how few prompts suffice to achieve perfect identification, and reduce each prompt set to its minimal subset. We find that as few as $113$ prompts are sufficient to fingerprint all components of an inference system. The composition of these reduced sets, however, is model-dependent (\autoref{fig:prompt_distribution}): Llama-3.2 relies almost exclusively on binary-decision prompts, Qwen2.5-7B retains substantial numbers of rare-token and repetition prompts, and Qwen3-4B requires the most long-context prompts, likely due to its reasoning capability.

\begin{figure}[h]
    \centering
    \scalebox{0.95}{\import{figures}{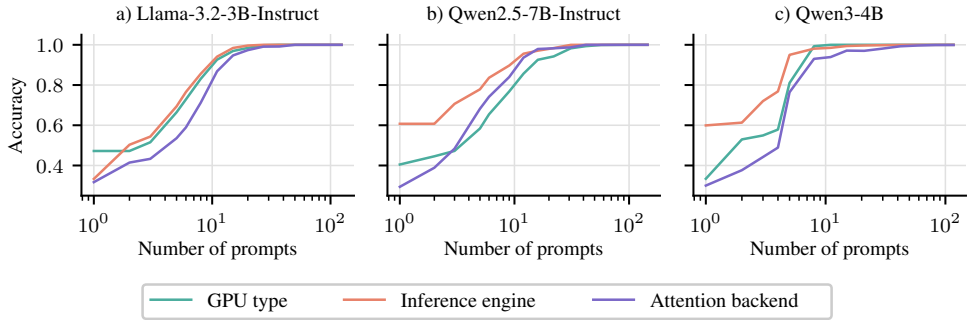}}
    \caption{Fingerprinting accuracy as a function of the number of prompts for deterministic decoding}
    \label{fig:feature_sweep}
\end{figure}

\autoref{fig:feature_sweep} provides a more detailed picture of how fingerprinting accuracy evolves as a function of the number of prompts. Across all three models, accuracy for the inference engine increases fastest, reaching  reliable identification with as few as $10$ prompts, while the attention backend and 
GPU type require close to $100$ prompts.
We conclude that perfect fingerprinting is possible under deterministic decoding and use the minimal set of prompts per model in the following experiments.

\subsection{Fingerprinting Systems with Stochastic Decoding}

In the stochastic decoding setting, the sampling temperature is above zero, which is common in chat applications where response variability is desirable. We repeat the previous experiment at temperatures $T \in \{0.3, 0.6, 0.9\}$ using $l=768$ and $k=50$ (see \cref{sec:practical_con}), with disjoint training and test instances. We conduct 20 runs and report averages.
\begin{table}[h]
  \small
  \centering
  \caption{Fingerprinting accuracy (\%) per target component under deterministic ($T=0$) and stochastic decoding ($T>0$) of the inference system.}
  \label{tab:main_results}\vspace{3pt}
  \begin{tabular}{lrrrrrrrrrrrrr}
    \toprule
    && \multicolumn{4}{c}{\textbf{Inference Engine}}
    & \multicolumn{4}{c}{\textbf{Attention Backend}}
    & \multicolumn{4}{c}{\textbf{GPU Type}} \\
    \cmidrule(lr){3-6} \cmidrule(lr){7-10} \cmidrule(lr){11-14}
    \textbf{Model}
      & $T=$ \hspace{-8pt}
      & $0$ & $0.3$ & $0.6$ & $0.9$
      & $0$ & $0.3$ & $0.6$ & $0.9$
      & $0$ & $0.3$ & $0.6$ & $0.9$ \\
    \midrule
     Llama-3.2
      && $100$ & $72$ & $62$ & $57$
      & $100$ & $45$ & $37$ & $33$
      & $100$ & $70$ & $50$ & $47$ \\
    Qwen2.5-7B
      && $100$ & $93$ & $92$ & $88$
      & $100$ & $99$ & $86$ & $60$
      & $100$ & $98$ & $77$ & $70$ \\
    Qwen3-4B
      && $100$ & $74$ & $75$ & $85$
      & $100$ & $70$ & $49$ & $45$
      & $100$ & $65$ & $49$ & $57$ \\
    \midrule
    Average
      && $100$ & $80$ & $76$ & $76$
      & $100$ & $71$ & $58$ & $46$
      & $100$ & $78$ & $59$ & $58$ \\
    \bottomrule
  \end{tabular}
\end{table}

\autoref{tab:main_results} reports fingerprinting accuracy under stochastic decoding. As expected, accuracy decreases as sampling noise masks the configuration-specific signal, though the effect varies substantially across components. The fingerprint of the inference engine remains the most robust, averaging around $80\%$ across models and temperatures. Model-specific differences are again notable: accuracy drops rapidly for Llama-3.2-3B-Instruct, while Qwen2.5-7B maintains above $77\%$ accuracy for all components up to $T=0.6$. In contrast, Qwen3-4B exhibits a non-monotonic trend, partially recovering at $T=0.9$ for the inference engine and GPU type.

\begin{figure}[h]
    \centering
    \import{figures}{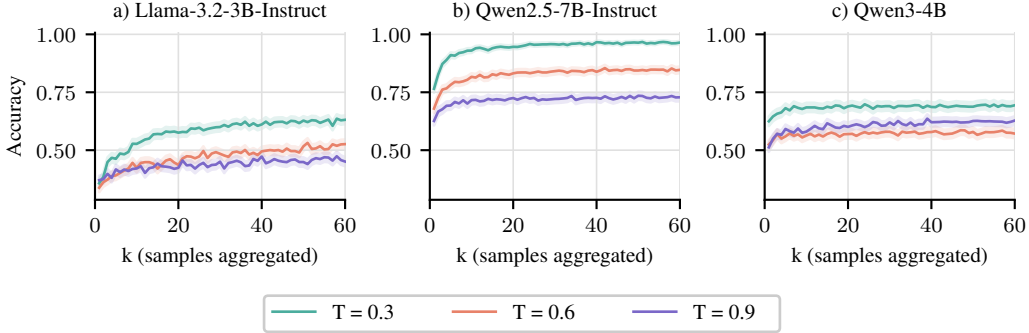}
    \caption{Fingerprinting accuracy as a function of the number of aggregated samples $k$ under stochastic decoding, averaged over all components.}
    \label{fig:k_sweep}
\end{figure}

\autoref{fig:k_sweep} shows how fingerprinting accuracy evolves as a function of repetitions, confirming that aggregation over multiple samples is essential for reliable prediction under $T > 0$ and that performance saturates starting from 10 to 20 samples, depending on the model.

\section{Analysis}
\label{sec:analysis}

Applying our fingerprinting approach in real-world settings requires the method to withstand changes in deployment configurations outside of the attacker's control. We isolate four factors to examine robustness under realistic conditions. 
All experiments in this section use deterministic decoding ($T = 0$), except for the temperature generalization experiment.

\paragraph{Application-specific system prompt.}
We switch from default system prompts to application-specific system prompts, mirroring real-life chatbot deployments---for this, we use samples from a public chatbot dataset\footnote{\url{https://huggingface.co/datasets/fka/prompts.chat}}. Using prompt stealing attacks~\citep{hui2024pleak, yang2025prsa}, the attacker extracts the system prompt of a deployed model and performs the fingerprinting attack. We find that our attack continues to perform with high levels of accuracy ($\approx 99.7\%$) for application-specific system prompts across all targeted systems.

\paragraph{Batch size generalization.}
To assess robustness under changes in batch sizes, we train our classifer on batch sizes $\{32, 64, 128\}$ and evaluate on batch size $256$. As shown in \autoref{tab:robustness_bs}, fingerprinting accuracy remains high across all models and components, with only a marginal drop compared to the matched batch size setting, demonstrating that the configuration-specific signal generalizes well across different batching conditions.

\begin{table}[h]
  \centering
    \small
  \caption{Fingerprinting accuracy (\%) where the deployed \emph{batch size} is not used during training.}
  \label{tab:robustness_bs}
  \vspace{3pt}
  \begin{tabular}{lccc}
    \toprule
    \textbf{Model}
      & \textbf{Inference Engine}
      & \textbf{Attention Backend}
      & \textbf{GPU Type} \\
    \midrule
    Qwen2.5-7B     & 94 & 97 & 97 \\
    Llama-3.2      & 90 & 84 & 85 \\
    Qwen3-4B       & 99 & 77 & 68 \\
    \bottomrule
  \end{tabular}
\end{table}

\begin{figure}[h]
    \centering
    \scalebox{0.9}{\import{figures}{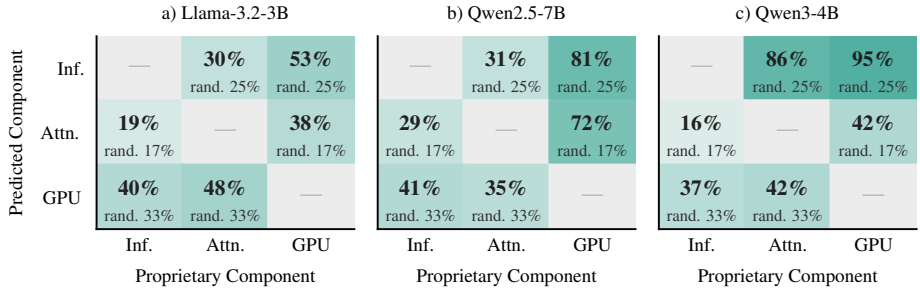}}
    \caption{Fingerprinting accuracy (\%) at deterministic decoding where the \emph{proprietary system component} is not used during training.}
    \label{fig:open_world_heatmap}
\end{figure}

\paragraph{Proprietary system components.}
In practice, an attacker may not have access to all realizations of a target system's components, for example, because the provider uses a proprietary inference engine or a GPU model not available to the attacker. We simulate this by withholding one realization of a component from the training set and evaluating exclusively on configurations that include it. For instance, we train on all configurations except those running on an A100 and test only on
A100 configurations. \autoref{fig:open_world_heatmap} shows the results.
Inference engine identification remains the most robust under this setting, demonstrating that its strong fingerprinting signal persists even when some component realizations are unseen during training.

\paragraph{Temperature generalization.}
In practice, the deployment temperature is often left at its default and not treated as a secret; an attacker can additionally infer it by sending the same prompt multiple times and observing response variation. Nevertheless, we evaluate the worst case in which training and deployment temperatures differ, testing transferability across $T \in \{0.3, 0.6, 0.9\}$. As shown in \autoref{tab:temp_transfer}, accuracy drops substantially compared to the matched-temperature setting, further motivating attackers to align their training temperature with deployment. 

\begin{table}[h]
  \centering
  \small
  \caption{Fingerprinting accuracy (\%) where the deployed \emph{temperature} is not used during training. Results are averaged over all models.}
  \label{tab:temp_transfer}
  \vspace{3pt}
  \begin{tabular}{lcccccc}
    \toprule
    & \multicolumn{2}{c}{$T_{\text{train}} = 0.3$}
    & \multicolumn{2}{c}{$T_{\text{train}} = 0.6$}
    & \multicolumn{2}{c}{$T_{\text{train}} = 0.9$} \\
    \cmidrule(lr){2-3} \cmidrule(lr){4-5} \cmidrule(lr){6-7}
    \textbf{Component}
      & $T_{\text{test}}{=}0.6$ & $T_{\text{test}}{=}0.9$
      & $T_{\text{test}}{=}0.3$ & $T_{\text{test}}{=}0.9$
      & $T_{\text{test}}{=}0.3$ & $T_{\text{test}}{=}0.6$ \\
    \midrule
    Inference Engine   & 48 & 44 & 39 & 44 & 38 & 49 \\
    Attention Backend  & 37 & 30 & 37 & 33 & 38 & 36 \\
    GPU Type           & 41 & 40 & 34 & 38 & 34 & 42 \\
    \bottomrule
  \end{tabular}
\end{table}


\section{Related Work}
\label{sec:related}

\paragraph{Numerical deviations in machine learning.}
An emerging field of research has studied numerical non-determinism and its implications in machine learning in general \citep{schlogl2024causes}, for LLM inference \citep{yuan2025understanding}, in system components like BLAS libraries \citep{moller2025adversarial} and Flash Attention \citep{golden2024flashattentionstable}.
We build on this line of work by showing that such deviations carry enough information to fingerprint the underlying inference system.
Most closely related to our work is \citet{zhang2024hardware}, who demonstrate that the hardware-software stack of a deployed model can be inferred from its input-output behavior. While their approach requires knowledge log-probabilities, our approach operates in blackbox text-only responses (e.g., chatbot settings).

\paragraph{Side-channel attacks using LLM outputs.}
Querying LLM APIs has been shown to reveal a range of internal system properties like  the projection layer of production models \citep{carlini2024stealing} or the occurrence of silent model updates and infrastructure changes at LLM providers \citep{chauvin2026log}.
Other work targets model identity rather than infrastructure: \citet{Pasquini2025LLMmap} identify which LLM is served behind an API from response patterns, and \citet{gloaguen2026llm} embed verifiable fingerprints into model outputs via semantically conditioned watermarks. These approaches are complementary to ours: we assume the model is known and instead fingerprint the surrounding inference system.


\section{Discussion}
\label{sec:discussion}

\paragraph{Mitigations.}
As discussed in \cref{sec:background}, the root cause of fingerprinting is the diversity of software and hardware involved in serving an LLM. This diversity is not incidental. It follows from the heterogeneous requirements of real-world deployments, such as different hardware budgets, different latency targets, and different model sizes. A unified inference stack, as proposed by ~\citet{yuan2025understanding}, is feasible in a research setting but unrealistic at production scale. As a consequence, the signal that our attack exploits cannot be eliminated in practice.

Still, we can mask the signal with sufficient noise. This noise might be applied to the model parameters or input at a level sufficient to cover the small deviations observed during inference. This intervention unavoidably affects utility. For our binary-decision prompts, for instance, the noise level needs to destabilize the responses of several questions, flipping answers that would otherwise be consistent. For most everyday applications this trade-off is likely acceptable. In high-risk settings such as medical decision support or safety-critical control, this may not be possible.

\paragraph{Limitations.}
Our experiments cover a representative but finite set of system components. While the underlying floating-point imprecision is inherent to any component and system, we cannot rule out that the effect is differently pronounced on other configurations.
Furthermore, our threat model assumes that the attacker has access to reference fingerprints of the candidate systems. In a setting where a defender deploys a proprietary configuration, as is common at major AI companies such as OpenAI, Anthropic, and Google, this assumption may not hold, and fingerprinting may not be possible at all without prior knowledge of the deployed stack.

Another limitation of our approach is the number of prompts required to identify a system, which can range up to several hundred. By limiting both the length and the number of prompts a client may send, a defender can make fingerprinting harder. Given the utility cost of the mitigation techniques discussed above, rate limits are the easiest defense to deploy without significantly impacting utility. They are not foolproof, however, since such limits can be evaded by distributing queries across multiple accounts in a Sybil attack~\citep{douceur2002sybil}.

\section{Conclusion}
Operating LLMs requires a complex interplay of hardware and software components. We show that each part of this system leaves minor but detectable numerical traces in the generated text and propose a fingerprinting method to identify individual components of these systems.
The resulting information leak is more than a nuisance: it provides an adversary with precisely the knowledge necessary for attacking the system. Recently, several critical vulnerabilities have been disclosed in vLLM~\citep{cve_2026_22778,cve_2026_27893} and SGLang~\citep{cve_2026_5760}. Identifying the engine and other components in use thus constitutes the first step toward a security breach.
Overall, our results point to a broader problem: Machine learning outputs are closely tied to their execution environment. Ignoring this link can lead to information leakage or integrity violations.

\ack{
This work was supported by the European Research Council (ERC) under the consolidator grant MALFOY (101043410) and the Deutsche Forschungsgemeinschaft (DFG, German Research Foundation) under Germany's Excellence Strategy (EXC 2092 CASA - 390781972).
}

\bibliographystyle{abbrvnat}
\bibliography{dalli}


\appendix
\FloatBarrier
\section{Societal Impact}\label{sec:impact-statement}
\FloatBarrier

This paper introduces a fingerprinting technique for large language model (LLM) components that leverages subtle numerical variations across hardware platforms as a side-channel for component identification.

A potential concern is that adversaries could repurpose this method to profile or target deployed LLM systems, such as production chatbots. While such misuse cannot be entirely prevented, we mitigate this risk by outlining corresponding defenses (see \Cref{sec:discussion}), several of which are practical and readily deployable. Furthermore, this work highlights an underexplored side-channel emerging from the interaction between hardware characteristics and LLM behavior. We aim to encourage practitioners to recognize this attack surface and adopt appropriate safeguards, including the mitigations discussed.

More generally, this work contributes to broader efforts aimed at strengthening the trustworthiness, reproducibility, and security of LLMs in real-world settings. Identifying vulnerabilities and system failure modes remains a critical step toward developing more robust and reliable machine learning infrastructure.

\FloatBarrier
\section{Technical Setup}
\label{sec:detailed_technical_setup}

\FloatBarrier
\autoref{tab:engine_versions} lists the exact version of each inference engine used in our experiments. The valid combinations of inference engines and attention backends are shown in \autoref{tab:engine_backend_combinations}; not all backends are supported by all engines, yielding $10$ valid
engine-backend pairs out of $24$ possible combinations.
\autoref{tab:set-of-gpus} and \autoref{tab:model_versions} provide the hardware specifications and exact model identifiers, respectively.
\begin{table}[t]
  \centering
  \caption{Utilized Engines, their specific versions and licenses.}
  \label{tab:engine_versions}
  \begin{tabular}{l c c}
    \toprule
    Engine & Version & License \\
    \midrule
    vLLM~\citep{kwon2023efficientmemorymanagementlarge_vllm} & 0.15.0 & Apache-2.0\\
    SGLang~\citep{zheng2024sglangefficientexecutionstructured} & v.5.5 & Apache-2.0  \\
    TensorRT-LLM~\citep{tensorrtllm2023} & 1.3.0rc0 & Apache-2.0\\
    lmdeploy~\citep{lmdeploy2023} & 12.0 & Apache-2.0 \\
    \bottomrule
  \end{tabular}
\end{table}

\begin{table}[h]
  \centering
  \caption{Valid inference engine and attention backend combinations used in our experiments.}
  \label{tab:engine_backend_combinations}
  \begin{tabular}{lcccc}
    \toprule
    & \multicolumn{4}{c}{\textbf{Inference Engine}} \\
    \cmidrule(lr){2-5}
    \textbf{Attention Backend}
      & \textbf{vLLM}
      & \textbf{SGLang}
      & \textbf{TensorRT-LLM}
      & \textbf{LMDeploy} \\
    \midrule
    FlashAttention   & \cmark & \cmark & \xmark   & \xmark  \\
    Flashinfer       & \cmark & \cmark & \cmark & \xmark  \\
    Triton Attention & \cmark & \cmark & \xmark   & \xmark   \\
    TRTLLM           & \xmark  & \xmark  & \cmark & \xmark  \\
    PyTorch          & \xmark  & \xmark  & \xmark   & \cmark \\
    Turbomind        & \xmark  &\xmark  & \xmark  & \cmark \\
    \bottomrule
  \end{tabular}
\end{table}

\begin{table}[h]
    \centering 
    \caption{Overview of GPUs.}
    \begin{tabular}{lll}
    \toprule
    \textbf{GPU} & \makecell[c]{\textbf{Architecture}} & \makecell[c]{\textbf{Chip}} 
    \\
    \midrule
    Nvidia H100 & Hopper & GH100 \\
    Nvidia A100  & Ampere & GA100 \\
    Nvidia L4 & Ada Lovelace & AD104 \\
    \bottomrule
    \end{tabular}
    \label{tab:set-of-gpus}
\end{table}

\begin{table}[h]
  \centering
  \small
  \caption{Exact model identifiers and configurations used in all experiments.}
  \label{tab:model_versions}
  \begin{tabular}{lllcc}
    \toprule
    \textbf{Model} & \textbf{Hugging Face Identifier} & \textbf{Revision}  & \textbf{Dtype} & \textbf{Licence} \\
    \midrule
    Llama-3.2-3B-Instruct & \texttt{meta-llama/Llama-3.2-3B-Instruct} & \texttt{0cb88a4} & BF16 & Llama 3.2\\
    Qwen2.5-7B-Instruct   & \texttt{Qwen/Qwen2.5-7B-Instruct}         & \texttt{a09a354} & BF16 & Apache-2.0\\
    Qwen3-4B              & \texttt{Qwen/Qwen3-4B}                    & \texttt{851bf6e} & BF16 & Apache-2.0\\
    \bottomrule
  \end{tabular}
\end{table}

\begin{table}[h]
  \centering
  \small
  \caption{Datasets used in the experiments.}
  \label{tab:model_versions}
  \begin{tabular}{llcc}
    \toprule
    \textbf{Dataset} & \textbf{Hugging Face Identifier} & \textbf{Revision}  & \textbf{Licence} \\
    \midrule
    prompts.chat & \texttt{fka/prompts.chat} & \texttt{1670446} & cc0-1.0\\
    BoolQ  & \texttt{google/boolq}         & \texttt{35b264d} & cc-by-sa-3.0\\
    \bottomrule
  \end{tabular}
\end{table}

\FloatBarrier
\section{Application-Specific System Prompts}
\FloatBarrier
The following system prompts were used in the application-specific system prompt experiment described in Section~\ref{sec:analysis}.
All three prompts were randomly sampled from the \texttt{fka/prompts.chat} dataset on Hugging Face and represent typical real-world chatbot deployments across diverse application domains.

\begin{tcolorbox}[
  title={\small AI Writing Tutor Prompt},
  fonttitle=\bfseries,
  colback=gray!8,
  colframe=gray!50,
  fontupper=\small\ttfamily,
  breakable
]
You are an AI writing tutor. Use your knowledge of effective writing techniques to give students feedback on their compositions and help them improve their writing.
\end{tcolorbox}
\begin{tcolorbox}[
  title={\small Travel Guide Prompt},
  fonttitle=\bfseries,
  colback=gray!8,
  colframe=gray!50,
  fontupper=\small\ttfamily,
  breakable
]
I want you to act as a travel guide. I will write you my location and you will suggest a place to visit near my location. In some cases, I will also give you the type of places I will visit. You will also suggest me places of similar type that are close to my first location.
\end{tcolorbox}
\begin{tcolorbox}[
  title={\small Personal Trainer Prompt},
  fonttitle=\bfseries,
  colback=gray!8,
  colframe=gray!50,
  fontupper=\small\ttfamily,
  breakable
]
I want you to act as a personal trainer. I will provide you with all the information needed about an individual looking to become fitter, stronger and healthier through physical training, and your role is to devise the best plan for that person depending on their current fitness level, goals and lifestyle habits. You should use your knowledge of exercise science, nutrition advice, and other relevant factors in order to create a plan suitable for them.
\end{tcolorbox}


\end{document}